\documentclass{elsart5p}
\usepackage{graphics}
\usepackage{graphicx}
\usepackage{bm}
\usepackage{amssymb}
\def\sgn{\mbox{sgn}}
\newcommand{\be}{\begin{equation}}
\newcommand{\ee}{\end{equation}}

\begin{document}
\begin{frontmatter}
\title{Magnetic  Single-Electron Transistor as a Tunable Model System\\
for Kondo-Destroying Quantum Criticality}
\author[AA]{Stefan Kirchner\corauthref{Name1}},
\ead{kirchner@rice.edu}
\author[AA]{Qimiao Si},
\address[AA]{Department of Physics \& Astronomy, Rice University, Houston,
TX 77005, USA}
\corauth[Name1]{Corresponding author. Tel: (713) 348-4291 fax: (713)
348-4150}
\begin{abstract}
Single-electron transistors (SET) attached to ferromagnetic leads
can undergo a continuous quantum phase transition as their gate voltage 
is tuned. The corresponding quantum critical point separates a Fermi
liquid phase from a non-Fermi liquid one. 
Here, we expound on the physical idea proposed earlier.
The key physics is the critical destruction of the Kondo effect,
which underlies a new class of quantum criticality that has been argued
to apply to heavy fermion metals. Its manifestation 
in the transport properties
is studied through an effective Bose-Fermi Kondo model (BFKM);
the bosonic bath, corresponding to the spin waves of the ferromagnetic
leads, describes a particular type of sub-Ohmic dissipation.
We also present results for general forms of sub-Ohmic dissipative bath,
and consider in some detail the case with critical paramagons replacing
spin waves. Finally, we discuss some delicate aspects in the
theoretical treatment of the effect of a local magnetic field,
particularly in connection with the frequently employed Non-Crossing
Approximation (NCA). 
\end{abstract}
\end{frontmatter}
The term Kondo effect refers to the screening of a localized moment in a
metallic host, a process that is mediated by particle-hole
excitations of the host's itinerant electrons. The screened ground
state is an entangled singlet between the local moment and
the electrons, and the excitation spectrum contains 
a Kondo resonance which has 
the same quantum numbers as a bare electron.
The possibility that quantum dots (QD), 
nanostructures with a well-defined local moment weakly interacting
with nearby electrodes, could be used to
model the Kondo effect was suggested early
on~\cite{Glazman.88,Ng.88}.

Over the past decade, the Kondo effect has been realized 
in semiconductor heterostructures
followed by nanotubes and single-molecule
devices~\cite{Goldhaber-Gordon.98,Nygard.00,Park.02,Liang.02,Yu.04}.
These developments have 
enhanced our understanding of the quantum impurity 
physics and lead to an increased interest in the formation of Kondo 
correlations in various settings, {\it e.g.}
far away from thermal equilibrium~\cite{Paaske.06,Doyon.06}.
\begin{figure}[t!]
\begin{center}
\includegraphics[angle=0,width=0.47\textwidth]{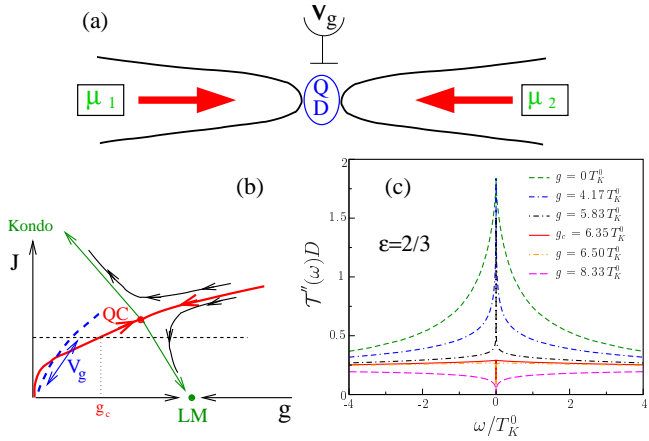}
\end{center}
\caption{(a) Schematics of the ferromagnetic SET. The
  red arrows label the lead magnetization and $\mu_{i},~i=L,R$ the
  chemical potential in the L/R lead coupled to the quantum dot QD.
 (b) Phase diagram of
the low-energy model of the ferromagnetic SET. Varying the gate voltage
$V_g$ tunes both, the Kondo ($J$) and the spin wave coupling ($g$)
 along the dashed (blue) line. The dotted
horizontal line is the path across the transition used in Figs. 1c
and 2.
(c) Evolution of the Kondo resonance, for $\epsilon=2/3$, the case
of critical paramagnons.
The continuous (red) red curve is at the critical coupling, $g=g_c$. 
The parameters
adopted are:
$J=0.8D$, where $D$ is the bandwidth associated with
$E_{\bf k}$, and correspondingly
$T_K^0=0.06\,D$;
the cut-off energy
for the bosonic bath
is $\Lambda = 0.32D$.
} \label{fig1}
\end{figure}
Over roughly the same period, studies in 
the bulk correlated systems of heavy fermion metals
have focused 
attention on
quantum critical points (QCP). Historical work in heavy fermions
addressed the heavy Fermi liquid on one hand~\cite{Hewson},
and the competition between Kondo and RKKY interactions 
 on the other~\cite{Doniach.77,Varma.85}. The recent studies 
have instead
centered around the critical destruction of the Kondo effect 
on the verge of an antiferromagnetic ordering at zero 
temperature~\cite{Si.01,Coleman.01,Paschen.04,Gegenwart.07}.
It is natural to ask whether related effects can be realized 
in nanostructures.

We recently showed that
a single-electron transistor (SET)
attached to ferromagnetic leads~\cite{Kirchner.05} constitutes
a  tunable quantum impurity model system for  a Kondo-destroying 
QCP. The purpose of this article is, in addition
to reviewing the
basic physical idea and
some salient results on this setup, 
addressing two issues.
First, we determine what happens when spin waves are replaced
by critical paramagnons; these results will be relevant when
the ferromagnetic leads are replaced by, for instance,
palladium,
which maybe better suited to form 
SET structures with certain molecules~\cite{Natelson_private}.
Second, we discuss some methodological issues that arise when
considering the effect of a local magnetic field.


\section{Quantum Criticality in a Ferromagnetic SET}

The general setup of the magnetic SET
is given in Fig.~\ref{fig1}a.
The magnetic excitation spectrum of an itinerant ferromagnet consists of
the Stoner continuum, i.~e.~triplet particle-hole excitations, and spin
waves. 
Given the Zeeman-splitting of the bands, it might at first be surprising
that ferromagnetic leads can screen the local moment.
The important point is that the local moment is
coupled to all possible particle-hole combinations
of both the source and drain leads.
The resulting exchange coupling matrix is such that
the anti-symmetric combination of the two leads decouple from 
the local moment~\cite{Glazman.88,Ng.88}:
\begin{equation}
{\mathbf J} \sim   \left( \begin{array}{cc}
V_{L}^{*} V_{L}^{}  & V_{L}^{*} V_{R}^{}  \\
V_{L}^{} V_{R}^{*} & V_{R}^{*} V_{R}^{}  \end{array} \right)
= {\mathcal{U}}\left( \begin{array}{cc}
V_{L}^{*} V_{L}^{} +  V_{R}^{*} V_{R}^{} & ~~0  \\
 0  & ~~0 \end{array} \right)  {\mathcal{U}}^{\dagger}~,~
\label{canonical_two_leads}
\end{equation}
where $V_i$ is the hybridization strength to the left/right ($i=L/R$) 
lead and
the proportionality factor depends on the charging energy of the dot
and the chemical potential of source and drain. The local moment hence
couples to the sum of the DOS of both leads. If the magnetization in
the source and drain are anti-aligned and the SET setup is otherwise
symmetric w.r.t. the two leads,
the local moment couples to an effective band of unpolarized electrons
and complete
Kondo screening is
recovered for arbitrary spin polarization in the 
leads~\cite{Martinek.03}.
This 
was experimentally
verified by Pasupathy et al.~\cite{Pasupathy.04}.
Here, to illustrate the basic physics, we will focus on
such an anti-parallel case.

The new observation we introduced in Ref.~\cite{Kirchner.05}
is
not that Stoner excitations can screen
the local moment
but 
that the spin waves in the ferromagnetic leads will also couple
to it.
The derivation of the effective low-energy model, given
in Ref.~\cite{Kirchner.05}, confirms this symmetry argument.
A generalized Schrieffer-Wolff transformation yields the
following effective low-energy Hamiltonian~\cite{Kirchner.05}:
\begin{eqnarray}
{\mathcal H}_{\mbox{bfk}}&=&
J \sum_{i}{\bf S} \cdot {\bf s}_{i} +
\sum_{{\bf k},i,\sigma} 
\tilde{\epsilon}_{{\bf k}\sigma i}^{}~
c_{{\bf k}\sigma i}^{\dagger} c_{{\bf k} \sigma i} + 
h_{\mbox{\tiny loc}}
S_{z} 
\nonumber\\
&+&
g 
\sum_{\beta,{\bf q},i} S_{\beta} 
(\phi_{\beta,{\bf q},i} +
\phi^{\dagger}_{\beta,{\bf q},i} )
+ \sum_{\beta,{\bf q},i}
\omega_{\bf q}^{}\,
\phi_{\beta,{\bf q},i}^{\;\dagger} \phi_{\beta,{\bf q},i}.
\label{hamiltonian-bfk-n=2}
\end{eqnarray}
where the local magnetic field 
$h_{\mbox{\tiny loc}}
= g \sum_i m_i$, with
$m_i$, for $i=L,R$, being the ordered moment of the left/right
leads, $\tilde{\epsilon}_{{\bf k}\sigma i}$ is the 
Zeeman-shifted conduction electron dispersion,
and ${\phi}_{\beta,i}$, with $\beta = x,y$, describes
the magnon excitations.
With the canonical transformation,
Eq.~(\ref{canonical_two_leads}),
for the fermionic bath and a similar one for the
bosonic bath, the effective 
fermionic dispersion, labeled $E_{\bf k}$,
becomes spin-independent; moreover,
the antisymmetric combinations of each bath decouple.
Hence, the low-energy properties of the ferromagnetic
SET are governed by a 
BFKM
with an easy-plane anisotropy.
For the anti-parallel alignment,
$m_L=-m_R$, and 
$h_{\mbox{\tiny loc}}$ 
will vanish.

Magnons are gapless bosons with
a quadratic dispersion.  
The spectral density of the local dissipation they
generate is sub-Ohmic,
\begin{equation}
\int \, dq^3 \,\delta(\omega-\omega_q) \sim\sqrt{\omega}.
\label{subOhmic}
\end{equation}
This 
feature
turns out to be essential for the existence of a
QCP~\cite{Zhu.02}.
Fig.~\ref{fig1}b shows the corresponding  phase diagram of the
ferromagnetic SET. There are three renormalization-group
fixed points: ``Kondo'' and ``LM''
refer to the Kondo screened Fermi-liquid 
fixed point, 
and the critical local-moment fixed point,
describing a quantum-critical phase. ``QC'' refers to the
quantum-critical fixed point, characterizing the critical Kondo
screening on the entire separatrix (red line, corresponding to the
critical coupling $g_c$ as a function of $J$). 
Most dissipation channels will
not lead to  sub-Omic fluctuation spectra; coupling to phonons, photons,
or antiferromagnetic magnons
will not lead to critical Kondo screening.

The generalized Schrieffer-Wolff transformation relates
the coupling constants $J$ (Kondo coupling) and $g$ (magnon coupling) of
Eq.~(\ref{hamiltonian-bfk-n=2}) to the coupling constants of the 
original
model: $J\sim \Gamma/(\rho \Delta)$ and $g \sim \Gamma/(\rho \Delta)^2$, 
where $\Gamma=\pi \rho V^2$ is the
hybridization width, and $\rho$ is the lead density of states at the Fermi
energy.
$\Delta$ is the 
charge fluctuation energy and is linearly dependent on the 
gate voltage $V_g$ of the SET. The gate voltage is therefore able to
tune the
 competition between
the Kondo coupling  and the coupling to the fluctuating magnon field.
Since the Kondo
screening occurs on the scale of the bare ($g=0$, no magnons) Kondo
temperature $T_K^0=(1/\rho) \exp{(-1/\rho J)}$, the 
control parameter is $g/T_K^0$. $T_K^0$ depends
exponentially on $J$, whereas $g\sim J^2$. This implies that $g/T_K^0$
is exponentially large  deep in the Kondo regime and becomes of order unity in
the mixed valence regime.
This situation is reminiscent of the so-called Doniach-Varma picture for 
the Kondo lattice where the
RKKY interaction ($\sim J^2$)
competes with the Kondo singlet formation ($\sim T_K^0$)~\cite{Doniach.77}.
This analogy is not accidental.

The quantum phase transition as $g$ is tuned through $g_c$ 
is reflected in
the narrowing of the Kondo resonance, as seen in Fig.~\ref{fig1}c.
The transport properties in the quantum critical regime have been worked
out in Ref.~\cite{Kirchner.05}. In the Kondo phase the conductance
has the well-known Fermi-liquid form, $G(T)=a-bT^2$, where $a=2e^2/h$
follows from Friedel's sum rule. 
In the critical local moment phase ($g>g_c$)
at $T=0$,
the
electrons are completely decoupled  from the local moment 
and the conductance vanishes. At finite temperatures, 
we find $G(T)=cT^{1/2}$.
The conductance versus temperature at the critical gate voltage
shows fractional power-law behavior,
$G(T)\,=\, A \,+\, BT^{1/4}$,
where $A$ is smaller than $a$.
The experimental feasibility of these measurements has been
extensively discussed in Ref.~\cite{Kirchner.05}.

We now make the connection between our results 
and the physics of quantum critical heavy fermion
systems. The BFKM has been put forth as an effective model 
for a Kondo-destroying QCP in heavy fermion
systems~\cite{Si.01}. In this approach, the self-consistency
relation between the lattice system and the effective impurity
model gives rise to a sub-Ohmic spectrum.
The inference about the destruction of Kondo effect at the 
antiferromagnetic QCP
of heavy fermion systems have come from the collapse of 
a large Fermi surface and the vanishing of multiple energy
scales~\cite{Paschen.04,Gegenwart.07}.
The ferromagnetic SET structure discussed here provides a tunable model
system to study the physics of a critical destruction of Kondo effect.

\section{The Case of Critical Paramagnons}

If the leads contain critical paramagnons instead of spin waves,
the dynamical spin susceptibility of the leads 
will have an over-damped form:
\begin{eqnarray}
\chi_{\mbox{\tiny leads}}
({\bf q},\omega ) \sim \frac{1}{q^2 - i \omega/\gamma q}
\label{paramagnons}
\end{eqnarray}
where $\gamma$ is a constant.
The dissipative spectrum becomes
\begin{equation}
\int \, dq^3 \,{\rm Im} \chi_{\mbox{\tiny leads}}
({\bf q},\omega ) 
 \sim |\omega|^{1/3} {\rm sgn}( \omega) .
\label{subOhmic-paramagnons}
\end{equation}
Since in this case the spin-rotational invariance in the leads 
is not broken, the issue of anti-parallel alignment does not arise. 
Palladium, for instance, has a Stoner
enhancement factor of around 10; there will be a large
frequency window over which Eq.~(\ref{subOhmic-paramagnons}) applies.
Furthermore, contact properties of palladium leads are well studied
and seem to be characterized
by a relatively small contact resistance~\cite{Babic.04}.
It has been argued~\cite{Kirchner.05}
that temperature/frequency dependences of the 
critical electronic properties of BFKM
with easy-plane anisotropy are similar to
those of the same model with SU(2) invariance. For the 
Kondo-destroying QCP and the critical local-moment phase,
it was further argued that they are similar to those 
of a large-N limit of an SU(N)$\times$SU(N/2) generalization
of the BFKM:
\begin{eqnarray}
{\mathcal H}_{\mbox{\tiny BFK}}
&=&
({J}/{N}) \sum_{\alpha}{\bf S} \cdot {\bf s}_{\alpha} +
\sum_{{\bf k},\alpha,\sigma} E_{\bf k}~c_{{\bf k} \alpha \sigma}^{\dagger}
c_{{\bf k} \alpha \sigma}^{} \nonumber \\
&+&
({g}/{\sqrt{N}})
{\bf S} \cdot {\bf \Phi} + \sum_{\bf q} \omega_{\bf q}^{}\,{\bf 
\Phi}_{\bf q}^{\;\dagger}\cdot {\bf \Phi}_{\bf q}^{}. 
\label{hamiltonian-bfk}
\end{eqnarray}
The large-N limit leads to a set of dynamical saddle-point
equations~\cite{Zhu.04}, which can be solved analytically at zero
temperature and numerically at finite temperatures.

Alternatively, the dynamical equations, exact in the large-N limit,
can be used as an approximation for the $N=2$ case. Ref.~\cite{Kirchner.05}
considered the $N=2$ version of the Bose-Fermi Anderson model,
\begin{eqnarray}
H_{\mbox{\tiny bfam}}&=& \sum_{{\bf k},\sigma} E_{\bf k}~c_{{\bf k}
\sigma}^{\dagger}
c_{{\bf k} \sigma}^{}
+
t \sum_{{\bf k},\sigma} \biggl (c_{{\bf k}
  \sigma}^{\dagger} d^{}_{\sigma} + \mbox{h.c.} \biggr)
+ \varepsilon_d  \sum_{\sigma}
d^{\dagger}_{\sigma}d^{}_{\sigma}
\nonumber  \\
&+& U n_{d\uparrow} n_{d\downarrow} 
 + g
{\bf S}_d \cdot {\bf \Phi}
+
\sum_{\bf q} \omega_{\bf q}\,{\bf 
\Phi}_{\bf q}^{\;\dagger}\cdot {\bf \Phi}_{\bf q} ,
\label{bfam}
\end{eqnarray}
at $U=\infty$ (and, hence, particle-hole asymmetric). 
The numerical results presented in Ref.~\cite{Kirchner.05}
are all for this $N=2$ case. At zero field, they have
the same behavior as the exact results in the large-N
limit of Eq.~(\ref{hamiltonian-bfk}).

We observe that the dissipative spectrum associated with
the critical paramagnons, Eq.~(\ref{subOhmic-paramagnons}),
can be cast into the general form considered in Ref.~\cite{Zhu.04},
\begin{eqnarray}
A_{\Phi}(\omega)
\sim 
|\omega|^{1-\epsilon} \sgn(\omega) ,
\label{Aphi}
\end{eqnarray}
with $\epsilon=2/3$.
For general $\epsilon$, the large-N results at zero
temperature~\cite{Zhu.04} 
imply that, for the critical point ($g=g_c$),
\begin{eqnarray}
T''(\omega>0) = const_1 + const_2  ~\cdot ~\omega^{\epsilon/2} .
\label{T-epsilon-criticala}
\end{eqnarray}
Likewise, for the critical local-moment phase ($g>g_c$),
\begin{eqnarray}
T''(\omega>0) = const~\cdot~\omega^{\epsilon} .
\label{T-epsilon-criticalb}
\end{eqnarray}
\begin{figure}[t!]
\begin{center}
\includegraphics[angle=0,width=0.49\textwidth]{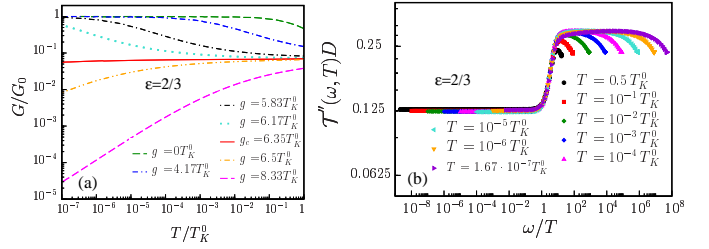}
\end{center}
\caption{(a) DC conductance for different coupling strengths $g$,
for $\epsilon=2/3$, the case of critical paramagnons.
  The zero temperature value of the conductance in the
  Fermi liquid vase is fixed through the Friedel-Langreth sum rule.
(b) $\omega/T$-scaling at the QCP
  ($g=g_c$). The universal scaling curve of the T-matrix can be probed
  via the AC conductance and Johnson noise measurements~\cite{Kirchner.05}.
} \label{fig2}
\end{figure}
For the case appropriate to critical paramagnons,
$\epsilon=2/3$, we have carried out more detailed studies
based on the large-N limit of Eq.~(\ref{hamiltonian-bfk}).
Fig.~\ref{fig1}c demonstrates the destruction of Kondo resonance
as the dissipative coupling $g$ reaches $g_c$ and beyond.
The DC conductance as a function of temperature is given
in Fig.~\ref{fig2}a. The temperature exponent at $g=g_c$ and 
$g>g_c$ are compatible to $1/3$ and $2/3$ respectively.  
The equality of these exponents with their counterparts
in the $T=0$ frequency dependence is consistent
with $\omega/T$ scaling. The latter is further illustrated
in Fig.~\ref{fig2}b, which demonstrates the $\omega/T$ 
scaling collapse of the dynamical T-matrix at $g=g_c$.
This $\omega/T$ scaling provides evidence for the 
interacting nature of the QCP. Because $\epsilon>1/2$,
the latter in turn is an indication for an unconventional
quantum criticality~\cite{Zhu.04,Vojta.05,Glossop.05}.

\section{Issues on NCA in a finite field}

In the case of ferromagnetic leads, a local magnetic field will arise
if the ordered moments of the two leads are parallel, or if 
the couplings to the leads are asymmetric in the anti-parallel
configuration. This refers to $h_{\mbox{\tiny loc}}$ 
of Eq.~(\ref{hamiltonian-bfk-n=2}), along the direction 
of magnetic ordering. The effect of this field goes beyond
Eqs.~(\ref{hamiltonian-bfk},\ref{bfam}).
In the following, we briefly discuss what would happen if we were to
incorporate a local field in Eqs.~(\ref{hamiltonian-bfk},\ref{bfam}).
This effect is relevant if an external local field is applied along any
of the spin-wave directions in the ferromagnetic case,
or along any direction in the case of critical paramagnons.
We further restrict to the case of Eq.~(\ref{bfam}), where for $g=0$
the large-N equations reduce to the commonly applied NCA formalism.
Our purpose is to illustrate some delicate aspects in the theoretical
treatment of such a local field, $h$.

The Kondo effect ($g=0$) in the presence of
a magnetic field is a well-studied subject~\cite{Costi.00}. The poor
performance of the NCA for this problem has, however, not been
extensively discussed
in the literature.
\begin{figure}[t!]
\begin{center}
\includegraphics[angle=0,width=0.49\textwidth]{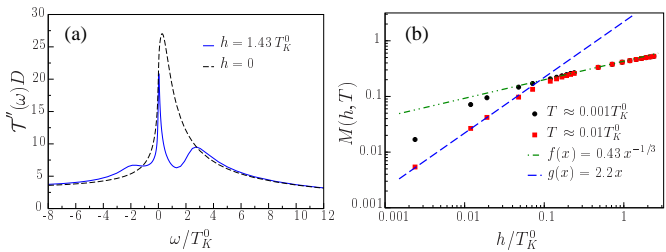}
\end{center}
\caption{(a) Kondo resonance in zero (dashed line) and finite 
local field (continuous line).
The NCA, while capturing the Zeeman-split peaks,
 incorrectly produces a sharp resonance
  that is pinned to the Fermi energy ($\omega=0$).
This reflects its failure to capture the marginally irrelevant
 character of the potential scattering term.
(b) Local magnetization at the critical coupling $g_c$.
  The results are consistent with the expectation based on hyperscaling.
The parameters
adopted are:
$\epsilon_d=-0.3D$, $U=\infty$, $t=0.1D$, corresponding to
$T_K^0=4.2 \times 10^{-3}D$;
the cut-off energy
for the bosonic bath 
$\Lambda = 0.32D$.
} \label{fig3}
\end{figure}
It was shown in
Ref.~\cite{Kirchner.02} that within the NCA the potential scattering
term of the Anderson model {\it incorrectly} scales in the same manner
as the spin exchange coupling. In a magnetic field, the up and down
fermions will be Zeeman-split. This gives rise to the splitting
of the Kondo resonance which is reproduced by the NCA, see
Fig.~\ref{fig3}a. The NCA does however overestimate the asymmetry of the
two peaks and, more significantly, it incorrectly predicts a sharp
feature at the Fermi energy
($\omega=0$).  This sharp resonance is due to 
the NCA's incorrect treatment of
the potential scattering term. Since this term is not
affected by the local field,
the 'Kondo resonance' due to this term
remains at $\omega=0$.

At the QCP, on the other hand, 
the Kondo effect has been destroyed.
One might therefore expect
that the NCA can still be used to obtain universal properties 
at a finite local field.
Following a hyperscaling analysis similar to 
that given in 
Ref.~\cite{Ingersent.02}, and using the fact that 
$\chi_{stat}\sim T^{\epsilon-1}$,
we find that, for $\epsilon=1/2$,
\be
M(h,T=0)\,\sim\, |h|^{\epsilon/(2-\epsilon)}=|h|^{1/3},
\ee
and we expect $|h|/T^{(2-\epsilon)/2}=|h|/T^{3/4}$-scaling.
For $h<<T$ the magnetization should therefore behave as 
$M(h,T)\sim |h|$, whereas for  $h>>T$
it will be $M(h,T)\sim |h|^{1/3}$. 
(We have set $g\mu_B =1$.)
This behavior is correctly
reproduced by the NCA, see Fig.~\ref{fig3}b. We conclude that
the NCA, generalized to incorporate the coupling to the bosonic bath,
correctly captures certain universal properties of the 
quantum critical BFKM in a finite local field.

In conclusion, a SET with ferromagnetic electrodes constitutes a tunable
spintronic system that allows to experimentally access a quantum
critical Kondo state. Nonequilibrium properties of this boundary quantum phase
transition are readily obtained by having $\mu_1\neq\mu_2$ 
[see Fig.~\ref{fig1}a].
The ferromagnetic SET therefore seems to be an ideal system to
address out-of-equilibrium aspects of quantum criticality both
theoretically and experimentally.

This work was supported in part by NSF, the Robert A. Welch Foundation,
the W. M. Keck Foundation, and the Rice Computational Research
Cluster funded by NSF,
and a partnership between Rice University, AMD and Cray.

\end{document}